# HDP: A Lightweight Cryptographic Protocol for Human Delegation Provenance in Agentic AI Systems

**Asiri Dalugoda**

Helixar Limited, Auckland, New Zealand

siri@helixar.ai | helixar.ai

*March 2026*

**Abstract**

Agentic AI systems increasingly execute consequential actions on behalf of human principals, delegating tasks through multi-step chains of autonomous agents. No existing standard addresses a fundamental accountability gap: verifying that terminal actions in a delegation chain were genuinely authorized by a human principal, through what chain of delegation, and under what scope. This paper presents the Human Delegation Provenance (HDP) protocol, a lightweight token-based scheme that cryptographically captures and verifies human authorization context in multi-agent systems. An HDP token binds a human authorization event to a session, records each agent's delegation action as a signed hop in an append-only chain, and enables any participant to verify the full provenance record using only the issuer's Ed25519 public key and the current session identifier. Verification is fully offline, requiring no registry lookups or third-party trust anchors. We situate HDP within the existing landscape of delegation protocols, identify its distinct design point relative to OAuth 2.0 Token Exchange (RFC 8693), JSON Web Tokens (RFC 7519), UCAN, and the Intent Provenance Protocol (draft-haberkamp-ipp-00), and demonstrate that existing standards fail to address the multi-hop, append-only, human-provenance requirements of agentic systems. HDP has been published as an IETF Internet-Draft (draft-helixar-hdp-agentic-delegation-00) and a reference TypeScript SDK is publicly available.

***Keywords:*** *agentic AI, delegation provenance, cryptographic authorization, multi-agent systems, prompt injection, Ed25519, human-in-the-loop security, IETF, authorization tokens.*

## 1. Introduction

The deployment of agentic AI systems has accelerated dramatically since 2023. Frameworks including LangChain [1], CrewAI [2], AutoGen [3], and OpenAI's Assistants API enable developers to compose pipelines in which a human instruction is passed to an orchestrator agent, which decomposes and delegates subtasks to specialized sub-agents, which in turn invoke tool-execution agents with access to file systems, databases, APIs, and external services. Production deployments of such systems now routinely execute thousands of consequential actions per day, including financial transactions, code commits, email dispatch, and data modification, with minimal human supervision at the action level.

This architecture creates a structural accountability gap. When a human authorizes an orchestrator, which delegates to sub-agents, which further delegate to tool-execution agents, the originating human authorization becomes progressively disconnected from terminal actions. There is currently no standard mechanism by which a downstream agent can verify that the action it is being instructed to perform was actually authorized by a human principal, under what scope that authorization was granted, and through what chain of delegation the instruction arrived.

This gap has three concrete operational consequences. First, post-hoc audits cannot reconstruct who approved what, and when. Second, agents have no runtime signal to distinguish a legitimate delegated instruction from an injected one, enabling prompt injection attacks [4, 5] to masquerade as authorized delegation. Third, accountability for autonomous actions cannot be attributed to a specific human authorization event, creating regulatory exposure as AI governance frameworks mature [6, 7].

We present the Human Delegation Provenance (HDP) protocol, a lightweight token-based scheme designed to close this gap. HDP defines a JSON token structure that: (1) records the human principal, their declared authorization scope, and a session binding at issuance; (2) accumulates a cryptographically signed hop record for each agent in the delegation chain; and (3) allows any recipient to verify the complete chain using only the issuer's Ed25519 public key and the current session identifier. Verification is fully offline, requiring no network call, registry lookup, or third-party trust anchor.

The remainder of this paper is organized as follows. Section 2 surveys related work and identifies the gap HDP occupies. Section 3 presents the threat model. Section 4 describes the protocol in full. Section 5 analyzes security properties. Section 6 discusses implementation considerations and reference artifacts. Section 7 presents limitations and future work. Section 8 concludes.

## 2. Related Work and Motivation

### 2.1 The Authorization Gap in Agentic Systems

The identity and access management challenges of agentic AI have attracted growing attention. The OpenID Foundation [8] has identified agentic delegation as a priority problem, noting that current protocols struggle to represent intricate sequences of delegations in which an agent may establish sub-agents or represent multiple principals concurrently. ISACA [9] characterizes this as a 'looming authorization crisis', observing that legacy IAM frameworks rely on preexisting scopes that are too coarse-grained and static to handle dynamic agent operational requirements. Strata [10] reports that non-human identities outnumber humans approximately 50:1 in the average enterprise, with 80% of IT leaders reporting agents acting outside expected behavior.

These analyses share a common diagnostic: existing authorization frameworks were designed for human-to-system interactions and do not capture the multi-hop, append-only, human-provenance requirements of agentic pipelines. HDP is designed explicitly to address this specific structural requirement.

**2.2 OAuth 2.0 Token Exchange (RFC 8693)**

RFC 8693 [11] defines a protocol for exchanging one security token for another, including delegation semantics. The specification defines an act claim for expressing that a principal is acting on behalf of another, and supports nesting of act claims to represent delegation chains. RFC 8693 is the closest existing standard to HDP's use case.

HDP and RFC 8693 are complementary rather than competing. RFC 8693 governs access token issuance within an OAuth 2.0 authorization server context, which requires a reachable token endpoint. HDP governs the provenance record that travels with an agentic task regardless of the authentication mechanism used. Three architectural differences are significant. First, RFC 8693 delegation is point-to-point, each exchange produces a new token. HDP carries an append-only chain within a single token, enabling the complete delegation history to be verified from a single artifact. Second, RFC 8693 requires authorization server availability. HDP verification is fully offline. Third, RFC 8693 makes no provision for human-specific authorization binding or scope provenance at the granularity required by agentic task authorization.

**2.3 JSON Web Token (RFC 7519) and JWT Bearer Tokens**

JSON Web Token [12] provides a general-purpose signed claims format widely used in agentic systems for agent-to-agent authentication. JWT-based approaches to agentic delegation, such as Agentic JWT [13], bind actions to authenticated intent and reduce escalation after injected instructions reach the action layer. HDP differs from JWT in three respects. First, HDP tokens carry an append-only, multi-party-signed delegation chain with no JWT equivalent. Second, HDP uses RFC 8785 canonical JSON for signing, rather than base64url-encoded header.payload. Third, HDP's verification pipeline is domain-specific to agentic delegation, incorporating session binding, hop verification, and max-hops enforcement.

**2.4 UCAN (User Controlled Authorization Networks)**

UCAN [14] defines a capability-based authorization token system using chained delegation with JWT. UCAN and HDP share the concept of delegation chains but differ significantly in scope and design goals. UCAN is a general capability authorization system; its delegation chain encodes executable capabilities enforced by receiving systems. HDP is specifically a provenance record for human-authorized agentic tasks: it makes no claims about capability enforcement and does not require the DID infrastructure that UCAN mandates. For deployments prioritizing offline operability and minimal infrastructure overhead, HDP presents a lower-complexity design point.

**2.5 Intent Provenance Protocol (IPP)**

The Intent Provenance Protocol [15] (draft-haberkamp-ipp-00) addresses the same root problem as HDP using Ed25519 signatures and append-only provenance chains. HDP and IPP are not interoperable and make different architectural trade-offs. Three differences are material to deployment context. First, revocation: IPP requires agents to poll a central revocation registry before every action; HDP uses short-lived tokens with session binding, requiring no registry at any point. Second, trust anchor: IPP tokens contain a genesis seal cryptographically linking every token to the specification author's public key at a specific URL, binding self-hosted deployments to a third-party key; HDP carries no genesis seal and imposes no external trust anchor. Third, identity model: IPP mandates W3C DID Core-conformant principal identifiers; HDP supports opaque identifiers as a first-class option.

### 2.6 Prompt Injection and the Accountability Surface

Prompt injection, where malicious content in an agent's environment overrides its intended behavior, is the primary practical attack class that HDP's provenance model addresses. Greshake et al. [4] demonstrated that indirect prompt injection enables data theft and service disruption in real-world LLM applications. Lee and Ryoo [5] showed that in multi-agent systems, injected prompts can self-replicate across agents in a worm-like propagation pattern. Ferrag et al. [16] present a unified taxonomy of 30+ attack techniques spanning input manipulation, model compromise, and protocol vulnerabilities, explicitly identifying cryptographic provenance tracking as a mitigation direction. The 2025 AI Agent Index [17] notes that only one of 30 surveyed deployed agent systems (ChatGPT Agent) implements any form of cryptographic request signing, and that this absence makes it significantly harder to prove what an agent actually did.

HDP does not prevent prompt injection at the semantic level. It provides an evidence trail that makes injected actions detectable in post-hoc audit: an injected action executed without a valid HDP token is attributable to a missing delegation event. An injected action recorded in the chain with a fabricated hop is detectable because fabrication requires the issuer's private key. This positions HDP as infrastructure for accountability, not a complete defense against injection.

### 2.7 Concurrent Work

Prakash [18] (arXiv:2603.24775) independently examines authorization provenance for agentic AI systems from a formal methods perspective. The threat model identified in that work is substantially consistent with the gap analysis in this paper, representing convergent recognition of the problem. The two works propose different technical approaches; HDP prioritizes offline verifiability and minimal infrastructure, while Prakash's framework emphasizes formal verification properties. We view these as complementary contributions to an emerging area.

## 3. Threat Model

HDP is designed to address a specific, bounded threat: the disconnection of terminal agent actions from their originating human authorization event in multi-agent pipelines. We define the threat model formally and state explicitly what HDP does and does not address.

### 3.1 Attacker Model

We consider an attacker who can: (A1) inject adversarial content into an agent's input stream through indirect prompt injection [4], including web pages, documents, database fields, and API responses; (A2) intercept and inspect tokens transmitted between agents; (A3) attempt to replay captured tokens in different sessions; (A4) attempt to forge or modify tokens to claim false authorization; (A5) attempt to insert fabricated hops into a delegation chain.

We do not model attackers who: possess the issuer's Ed25519 private key (key compromise is an operational security problem, not a protocol problem); can break Ed25519 signature security or SHA-512 collision resistance; or can tamper with the session establishment channel (assumed to be secured by transport-layer mechanisms).

**3.2 Assets and Goals**

The assets HDP protects are: (P1) the integrity of the human authorization event, including the principal identity, declared scope, and session binding; (P2) the integrity of the delegation chain, ensuring that recorded hops accurately reflect the actual delegation history; (P3) the non-replayability of tokens across sessions and after expiry.

HDP does not protect: agent behavior beyond what is recorded in the token; semantic correctness of actions relative to declared scope (this is an application layer concern); confidentiality of token contents (tokens are signed, not encrypted).

**3.3 Primary Attack Scenarios**

The four primary scenarios HDP is designed to detect are: (S1) an agent executing an action with no valid HDP token, indicating an unauthorized or injected instruction; (S2) a forged token presenting a false human principal or scope, detectable via root signature verification; (S3) a tampered chain in which a hop has been modified or removed, detectable via hop signature verification; (S4) a replayed token from a prior session, blocked by session binding.

## 4. Protocol Specification

**4.1 Design Principles**

HDP is designed according to five ordered principles. First, offline verifiability: verification requires only a public key and session ID. No network call, registry lookup, or third-party endpoint is required at any step. This enables use in air-gapped environments, edge deployments with intermittent connectivity, and latency-sensitive pipelines. Second, self-sovereignty: any organization can issue and verify HDP tokens without registering with a central authority or anchoring to a third-party key. Third, tamper evidence: any modification to any field, from header to any hop, is detectable by the verification pipeline. Fourth, minimal footprint: the protocol is implementable in any language with Ed25519 and JSON support. Fifth, privacy by design: principal identity fields are structurally separable from audit-relevant fields.

### 4.2 Token Structure

An HDP token is a JSON object with six top-level fields: the protocol version string (hdp), a header carrying lifecycle and session binding, a principal object identifying the authorizing human, a scope object recording the authorized intent and constraints, an append-only chain array of delegation hops, and a signature object carrying the root Ed25519 signature. Figure 1 shows the top-level structure.

```
{
"hdp" : "0.1",
"header" : { token_id, issued_at, expires_at, session_id, version },
"principal" : { id, id_type, display_name, poh_credential },
"scope" : { intent, authorized_tools, data_classification,
network_egress, persistence, max_hops },
"chain" : [ hop_1, hop_2, ... hop_n ],
"signature" : { kid, alg, value }
}
```

*Figure 1: HDP token top-level structure.*

#### 4.2.1 Header

The header carries a UUID v4 token identifier (token_id), Unix millisecond timestamps for issuance (issued_at) and expiry (expires_at, default 24 hours), a session identifier (session_id) established out-of-band between issuer and agent framework before token issuance, a version string mirroring the top-level hdp field, and an optional parent_token_id for re-authorization chaining.

#### 4.2.2 Principal

The principal object identifies the authorizing human with a required id and id_type. Supported id_type values are: opaque (application-defined, no resolution semantics), email, uuid, did (W3C DID [19]), and poh (Proof-of-Humanity credential). HDP mandates no specific identity model. The did type is available for deployments with existing DID infrastructure but is not required. Optional fields include display_name and poh_credential.

#### 4.2.3 Scope

The scope object records what the human authorized, is covered by the root signature, and must not be modified after issuance. Required fields are: intent (free-form natural language authorization statement), data_classification (one of: public, internal, confidential, restricted), network_egress (boolean), and persistence (boolean). Optional fields include authorized_tools, authorized_resources, and max_hops. Semantic validation of agent actions against declared scope is an application-layer concern; HDP provides the record, not the enforcement.

#### 4.2.4 Chain

The chain array is append-only. Each hop records: a sequential index (seq, starting at 1), agent identifier and type, an optional agent fingerprint, a Unix millisecond timestamp, a human-readable action summary, a parent hop index (0 for root human authorization), and a hop_signature. Agents must not remove or modify existing entries. Gaps in seq are a protocol violation.

### 4.3 Cryptographic Construction

HDP uses Ed25519 [20] (RFC 8032) for all signatures, with RFC 8785 JSON Canonicalization Scheme [21] providing deterministic serialization. Base64url encoding (RFC 4648 [22], no padding) is used for all binary fields.

*4.3.1 Root Signature*

The root signature is computed at token creation over the header, principal, scope, and empty chain. The procedure is: (1) construct the unsigned token object; (2) serialize to canonical JSON per RFC 8785, excluding the signature field; (3) compute Ed25519 signature over the canonical bytes; (4) base64url-encode and attach as signature.value. The signed payload is deterministically recoverable by any verifier.

*4.3.2 Hop Signatures*

Each hop carries a hop_signature binding the new hop record to the entire accumulated delegation history and to the root signature. The signing payload is constructed as a JSON array: [root_sig_value, hop_1, ..., hop_(n-1), new_hop_unsigned], where previously-signed hops are included WITH their hop_signature fields, and the new hop is included WITHOUT. This asymmetry is critical: the verifier must reconstruct this exact payload structure. Serialized per RFC 8785 and signed with the extending agent's private key. In HDP v0.1, all signatures use the issuer's key; multi-key delegation is a planned v0.2 feature.

The chaining construction means that every hop signature covers all prior hops and the root signature. Retroactive modification of any hop, or insertion of a fabricated hop, causes verification to fail for the tampered hop and all subsequent hops.

## 4.4 Verification Pipeline

A verifier must execute seven ordered steps. Failure at any step causes immediate rejection. The steps are: (1) version check; (2) expiry check against current time; (3) root signature verification using the issuer's public key; (4) hop sequence integrity check for gaps or duplicates; (5) hop signature verification for each hop using the payload reconstruction procedure of Section 4.3.2; (6) max_hops check against chain length; (7) session binding check comparing header.session_id to the verifier's current session. An optional eighth step validates a Proof-of-Humanity credential if configured.

The complete trust state required for verification is: the issuer's Ed25519 public key (32 bytes), the current session identifier, and the current time. No network calls are required at any step. This is a strong architectural guarantee enabling use in air-gapped, edge, and latency-sensitive environments.

## 4.5 Re-Authorization and Multi-Principal Delegation

Long-running sessions may exhaust max_hops, require scope expansion, or trigger re-authorization for high-risk actions. A new token supersedes the original by setting header.parent_token_id to the prior token's token_id before signing. The parentage link is cryptographically covered by the new root signature, creating an auditable lineage of scope evolution.

Multi-principal joint authorization is achieved by sequential chaining: Human A issues token T1; Human B issues token T2 with parent_token_id equal to T1. Each token is independently signed. Verification requires verifying each token individually, verifying parent_token_id linkage, and verifying shared session_id. This provides joint authorization auditably without threshold signature schemes, with each principal's authorization as a distinct signed artifact.

### 4.6 Transport

HDP tokens may be transmitted via the X-HDP-Token HTTP header (base64url-encoded JSON), or by reference via X-HDP-Token-Ref (UUID token_id with server-side storage). Tokens must not be transmitted in URL query parameters. Issuers may publish Ed25519 public keys at /.well-known/hdp-keys.json. IANA registration of both header fields and the application/hdp-token+json media type is requested in the IETF I-D.

## 5. Security Analysis

### 5.1 Token Forgery

A forged token, one whose header, principal, or scope fields differ from the original issuance, will fail step 3 of the verification pipeline. The security of this step reduces to the unforgeability of Ed25519 under chosen-message attack (EUF-CMA). Ed25519, specified in RFC 8032, is instantiated over Curve25519 with SHA-512 as the hash function. The EUF-CMA security of Ed25519 in the random oracle model follows from the hardness of the discrete logarithm problem over Curve25519 [23]. An attacker without the issuer's private key cannot produce a valid root signature for any modified token.

### 5.2 Chain Tampering

Modification of any hop, including field modification, removal, reordering, or insertion of a fabricated hop, causes hop signature verification to fail for the tampered hop and all subsequent hops. This follows from the chained construction of Section 4.3.2: each hop signature covers all prior hops with their signatures. The security reduces to EUF-CMA of Ed25519, with the same argument as Section 5.1. An attacker cannot fabricate a valid hop signature without the issuer's private key.

### 5.3 Replay Attack Defense

HDP provides two orthogonal replay defenses. Temporal expiry (expires_at, 24 hour default) ensures that captured tokens become invalid after a bounded window. Session binding (session_id) ensures that a token is valid only within the specific session for which it was issued. Together, these defenses ensure that a stolen non-expired token is useful to an attacker only within the original session. Applications with elevated security requirements should use short token lifetimes on the order of minutes rather than hours.

### 5.4 Prompt Injection Mitigation Boundary

HDP mitigates but does not fully prevent prompt injection (Section 3.1, attacker A1). An injected action that causes an agent to act without extending the HDP chain produces a detectable gap: a terminal action without a corresponding delegation record. An injected action recorded in the chain with a fabricated hop is detectable because fabrication requires the issuer's key. However, a sophisticated injected instruction that causes a legitimate agent to record a genuine hop with an action_summary misrepresenting the actual intent is not detectable by the protocol alone. This semantic validation boundary is an application-layer responsibility.

### 5.5 Privacy Properties

The principal object may contain PII. Issuers should use opaque identifiers and omit display_name when the receiving agent does not require human-readable identity. The structural separation of principal from audit-relevant fields enables implementations to strip principal from forwarded tokens; stripped tokens must be marked audit-only and must not be presented for signature verification, as principal removal invalidates the root signature. HDP tokens may constitute personal data under GDPR Article 4(1) where principal.id contains identifying information; implementations should apply appropriate retention controls.

### 5.6 Key Management Requirements

All HDP security guarantees depend on confidentiality of the issuer's Ed25519 private key. Implementations must store private keys in a secrets manager, HSM, or equivalent secure enclave. Private keys must not be stored in source code, configuration files, or environment variables in production. Key rotation is supported by issuing new tokens with a new kid value while maintaining the old key in the verifier's key set until all tokens signed with it have expired.

## 6. Implementation and Deployment

### 6.1 Reference Implementation

A reference implementation of HDP v0.1 is available as a TypeScript SDK (@helixar_ai/hdp) published on npm. The SDK provides token issuance, hop extension, and the seven-step verification pipeline. Python integrations for CrewAI and MCP are also available. The implementation uses the noble-ed25519 library for Ed25519 operations and a deterministic JSON canonicalization implementation conformant with RFC 8785.

### 6.2 Integration Patterns

HDP integrates with agentic orchestration frameworks at the orchestration layer. The typical integration pattern is: (1) human principal issues token via issuer SDK at session initiation; (2) token is attached to task context passed to the orchestrator; (3) each agent in the pipeline verifies the token on receipt, extends the chain with its intended action summary, and passes the extended token to downstream agents; (4) terminal tool-execution agents verify the complete chain before executing actions. Verification at every hop is recommended. Verification only at terminal agents is acceptable for performance-constrained deployments.

### 6.3 Performance Characteristics

Ed25519 signature verification is computationally inexpensive, completing in under 100 microseconds on contemporary hardware [24]. RFC 8785 JSON canonicalization is O(n) in token size. For a typical 10-hop delegation chain, full verification completes in under 2 milliseconds, making HDP suitable for high-throughput agentic pipelines. Token size grows linearly with chain length; a 10-hop token with typical field values is approximately 4-8 KB.

**6.4 IETF Internet-Draft**

The HDP protocol has been submitted as an IETF Internet-Draft (draft-helixar-hdp-agentic-delegation-00) to the RATS (Remote ATtestation procedureS) Working Group. The I-D defines the normative protocol specification, IANA considerations for HTTP header field registration and media type registration, and the comparison with related work surveyed in Section 2 of this paper. The draft is active as of March 2026 and is available at the IETF Datatracker.

## 7. Limitations and Future Work

**7.1 Single-Key Signing in v0.1**

HDP v0.1 uses the issuer's key for all hop signatures, meaning agents do not sign with their own keys. This simplifies key management but means hop signatures attest that a hop was recorded at the issuer, not that the specific agent produced it. HDP v0.2 will introduce per-agent key binding, enabling hop signatures that attest to the specific agent's identity using threshold or multi-signature schemes.

**7.2 Semantic Scope Enforcement**

HDP records what the human authorized but does not enforce it. Whether an agent's action is semantically consistent with scope.intent is an application-layer concern. Integration with runtime policy enforcement systems, such as MI9 [25] or CaMeL [26], is a natural extension: HDP provides the provenance record, and runtime enforcement systems use it as audit input to detect scope violations.

**7.3 Multi-Principal Simultaneous Authorization**

The current multi-principal model uses sequential chaining, which requires principals to act in sequence. A planned v0.2 extension introduces simultaneous multi-signature primitives using threshold signature schemes, enabling M-of-N human authorization without sequential dependency.

**7.4 Standardization Path**

The IETF I-D is currently individual submission status. Progressing to Working Group adoption in RATS or a related WG requires community review and demonstrated implementation. Feedback from deployment experience will inform future revisions. Alignment with emerging OpenID Foundation agentic identity work [8] is a priority for interoperability.

## 8. Conclusion

We have presented the Human Delegation Provenance (HDP) protocol, addressing a structural accountability gap in multi-agent AI systems: the disconnect of terminal actions from their originating human authorization. HDP provides a lightweight, offline-verifiable, self-sovereign token scheme that cryptographically binds human authorization events to agentic delegation chains. We have situated HDP within the existing landscape of delegation and authorization protocols, demonstrated that existing standards do not address the multi-hop, append-only, human-provenance requirements of agentic pipelines, and analyzed HDP's security properties against the relevant threat model.

The increasing capability and deployment of agentic AI systems makes the accountability gap HDP addresses a growing operational and regulatory risk. As Ferrag et al. [16] note, cryptographic provenance tracking is an identified mitigation direction for the class of protocol vulnerabilities emerging in LLM-agent ecosystems. The 2025 AI Agent Index [17] documents that only one of thirty surveyed deployed agent systems implements cryptographic request signing. HDP provides a minimal, deployable foundation for this capability.

The HDP specification is published as an IETF Internet-Draft. A reference TypeScript SDK and Python integrations are publicly available. Community feedback on the protocol design, IETF Working Group engagement, and deployment experience reports are welcomed.

## References

[1] Chase, H. et al. LangChain: Building applications with LLMs through composability. GitHub Repository. https://github.com/langchain-ai/langchain. 2022.

[2] CrewAI. Multi-agent orchestration framework. https://github.com/crewAIInc/crewAI. 2024.

[3] Wu, Q. et al. AutoGen: Enabling next-gen LLM applications via multi-agent conversation. arXiv:2308.08155. 2023.

[4] Greshake, K. et al. Not what you've signed up for: Compromising real-world LLM-integrated applications with indirect prompt injection. Proceedings of the 16th ACM Workshop on Artificial Intelligence and Security. 2023.

[5] Lee, D. and Ryoo, M. Prompt Infection: LLM-to-LLM prompt injection within multi-agent systems. arXiv:2410.07283. 2024.

[6] EU Artificial Intelligence Act. Regulation (EU) 2024/1689. Official Journal of the European Union. 2024.

[7] NIST. Artificial Intelligence Risk Management Framework (AI RMF 1.0). National Institute of Standards and Technology. 2023.

[8] OpenID Foundation. Identity Management for Agentic AI: The new frontier of authorization, authentication, and security for an AI agent world. OpenID Foundation Whitepaper. 2025.

[9] ISACA. The Looming Authorization Crisis: Why Traditional IAM Fails Agentic AI. ISACA Industry News. 2025.

[10] Strata Identity. Agentic AI Security: A Guide to Strategies for AI Agent Security. https://www.strata.io. 2026.

[11] Jones, M., Nadalin, A., Campbell, B., Bradley, J., and Liu, C. OAuth 2.0 Token Exchange. RFC 8693. IETF. January 2020. DOI: 10.17487/RFC8693.

[12] Jones, M., Bradley, J., and Sakimura, N. JSON Web Token (JWT). RFC 7519. IETF. May 2015. DOI: 10.17487/RFC7519.

[13] Goswami, A. Agentic JWT: Binding agent actions to authenticated intent. In: Survey of Agentic AI and Cybersecurity. arXiv:2601.05293. 2026.

[14] UCAN Working Group. User Controlled Authorization Network (UCAN) Specification v1.0. https://github.com/ucan-wg/spec. 2024.

[15] Haberkamp, M. Intent Provenance Protocol (IPP). Internet-Draft draft-haberkamp-ipp-00. IETF. 2024.

[16] Ferrag, M.A. et al. From Prompt Injections to Protocol Exploits: Threats in LLM-Powered AI Agents Workflows. arXiv:2506.23260. 2025.

[17] Casper, S. et al. The 2025 AI Agent Index: Documenting Technical and Safety Features of Deployed Agentic AI Systems. arXiv:2602.17753. 2026.

[18] Prakash, S. Authorization Provenance in Agentic AI Systems. arXiv:2603.24775. 2026.

[19] Sporny, M., Longley, D., Sabadello, M., Reed, D., Steele, O., and Allen, C. Decentralized Identifiers (DIDs) v1.0. W3C Recommendation. July 2022.

[20] Josefsson, S. and Liusvaara, I. Edwards-Curve Digital Signature Algorithm (EdDSA). RFC 8032. IETF. January 2017. DOI: 10.17487/RFC8032.

[21] Rundgren, A., Jordan, B., and Erdtman, S. JSON Canonicalization Scheme (JCS). RFC 8785. IETF. June 2020. DOI: 10.17487/RFC8785.

[22] Josefsson, S. The Base16, Base32, and Base64 Data Encodings. RFC 4648. IETF. October 2006. DOI: 10.17487/RFC4648.

[23] Bernstein, D.J. and Lange, T. SafeCurves: Choosing safe curves for elliptic-curve cryptography. https://safecurves.cr.yp.to. 2014.

[24] Bernstein, D.J. et al. Ed25519: High-speed high-security signatures. Journal of Cryptographic Engineering 2(2). 2012.

[25] MI9 Project. Agent Intelligence Protocol: Runtime Governance for Agentic AI Systems. arXiv:2508.03858. 2025.

[26] Debenedetti, E. et al. CaMeL: Defeating Prompt Injections by Design. arXiv preprint. 2025.

[27] Bradner, S. Key words for use in RFCs to Indicate Requirement Levels. RFC 2119. IETF. March 1997.

[28] Helixar Limited. HDP TypeScript Reference Implementation (@helixar_ai/hdp). npm. https://www.npmjs.com/package/@helixar_ai/hdp. 2026.

[29] Helixar Limited. Human Delegation Provenance Protocol (HDP) v0.1 Specification. https://helixar.ai/labs/hdp. 2026.

---

**Licence**

---